# Diversifying Citation Recommendations

ONUR KÜÇÜKTUNÇ, ERIK SAULE, KAMER KAYA and ÜMİT V. ÇATALYÜREK,
The Ohio State University

Literature search is arguably one of the most important phases of the academic and non-academic research. The increase in the number of published papers each year makes manual search inefficient and furthermore insufficient. Hence, automatized methods such as search engines have been of interest in the last thirty years. Unfortunately, these traditional engines use keyword-based approaches to solve the search problem, but these approaches are prone to ambiguity and synonymy. On the other hand, bibliographic search techniques based only on the citation information are not prone to these problems since they do not consider textual similarity. For many particular research areas and topics, the amount of knowledge to humankind is immense, and obtaining the desired information is as hard as looking for a needle in a haystack. Furthermore, sometimes, what we are looking for is a set of documents where each one is different than the others, but at the same time, as a whole we want them to cover all the important parts of the literature relevant to our search. This paper targets the problem of result diversification in citation-based bibliographic search. It surveys a set of techniques which aim to find a set of papers with satisfactory quality and diversity. We enhance these algorithms with a direction-awareness functionality to allow the users to reach either old, well-cited, well-known research papers or recent, less-known ones. We also propose a set of novel techniques for a better diversification of the results. All the techniques considered are compared by performing a rigorous experimentation. The results show that some of the proposed techniques are very successful in practice while performing a search in a bibliographic database.

Categories and Subject Descriptors: H.3.3 [**Information Storage and Retrieval**]: Information Search and Retrieval

This work was supported in parts by the DOE grant DE-FC02-06ER2775 and by the NSF grants CNS-0643969, OCI-0904809, and OCI-0904802.
Author's addresses: O. Küçüktunç, Department of Computer Science and Engineering, The Ohio State University; email: kucuktunc.1@osu.edu; E. Saule and K. Kaya and Ü. V. Çatalyürek, Department of Biomedical Informatics, The Ohio State University; emails: {esaule, kamer, umit}@bmi.osu.edu.

# 1. INTRODUCTION

The academic community has published millions of research papers to date, and the number of new papers has been increasing with time. For example, based on DBLP[1], computer scientists published 3 times more papers in 2010 than in 2000 (see Figure 1-left). With more than one hundred thousand new papers each year, performing a complete literature search became a herculean task. A paper cites 20 other papers on average (see Figure 1-right for the distribution of citations in our data), which means that there might be more than a thousand papers that cite or are cited by the papers referenced in a research article. Researchers typically rely on manual methods to discover new research such as keyword-based search via search engines, reading proceedings of conferences, browsing publication list of known experts or checking the reference list of the paper they are interested. These techniques are time-consuming and only allow to reach a limited set of documents in a reasonable time. Developing tools that help researchers to find relevant papers they do not know has been of interest for the last thirty years.

Some of the existing approaches and tools for the literature search cannot compete with some characteristics of today's literature. For example, keyword-based approaches suffer from the confusion induced by different names of identical concepts in different fields. (For instance, *partially ordered set* or *poset* are also often called *directed acyclic graph* or *DAG*). Conversely, two different concepts may have the same name in different fields (for instance, *hybrid* is commonly used to specify software hybridization, hardware hybridization, or algorithmic hybridization). These two problems may drastically increase the number of suggested but unrelated papers.

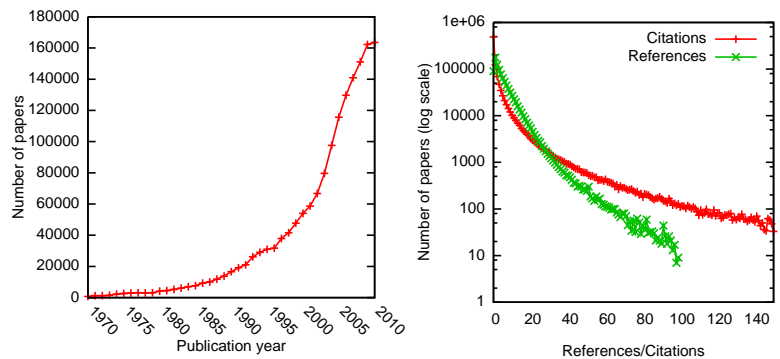

Fig. 1.   Number of new papers published each year based on DBLP (left), and number of papers with given citation and reference count (right).

Since they do not use textual information, bibliographic search techniques based only on the citation information do not suffer from above-mentioned problems [Kessler 1963; McNee et al. 2002; Small 1973; Lawrence et al. 1999; Liang et al. 2011; Gori and Pucci 2006; Lao and Cohen 2010; Li and Willett 2009; Ma et al. 2008]. Furthermore, it has been shown that text-based similarity is not sufficient for this task and that most of the relevant information is contained within the citation information [Strohman et al. 2007]. Besides, it is plausible that there is already a correlation between citation similarities and text similarities of the papers [Salton 1963; Peters et al. 1995] Following the idea of using citation information for bibliographic search, we built an efficient

---

[1] http://www.informatik.uni-trier.de/~ley/db/ statistics based on data acquired in Dec'11

and effective web service called the **advisor**[2] [Kucuktunc et al. 2012b; Kucuktunc et al. 2012a] based on personalized PageRank. It takes a bibliography file containing a set of papers, i.e., *seeds*, as an input to initiate the search. The algorithms employed by the **advisor** have the *direction-awareness* functionality which allow the user to specify her interest in classical or recent papers. Taking this criteria into account, the service returns a set of suggested papers ordered with respect to a ranking function. After obtaining the results, the user can give positive feedback to the system, and if desired, the output set is refined.

Today, for many particular research areas and topics, the amount of knowledge to humankind is immense and reaching the correct information is as hard as looking for a needle in a haystack. Furthermore, sometimes, what we are looking for is a set of results where each one is different than the others, but at the same time, as a whole we want them to cover all the important parts of the literature relevant to our search. Hence, diversifying the results of the search process is an important task to increase the amount of information one can reach via an automized search tool. There exist many recommender systems, such as Google web search, which personalize the output with respect to user query/history. It is a well known fact that for several applications, *personalization* can be an important limitation while reaching all the relevant information [Drosou and Pitoura 2010], and *diversification* can be used to increase the coverage of the results and hence, user satisfaction [Agrawal et al. 2009; Clarke et al. 2008; Mei et al. 2010; Gollapudi and Sharma 2009].

In this work, we target the bibliographic search problem and diversifying the results of the citation/paper recommendation process with the following objectives in mind: (1) the direction awareness property is kept, (2) the method should be efficient enough to be computable in real time, and (3) the results are relevant to the query and also diverse within the set. The contribution of this work is three-fold:

— We survey various random walk-based diversity methods (i.e., GRASSHOPPER [Zhu et al. 2007], DIVRANK [Mei et al. 2010] variants, and DRAGON [Tong et al. 2011]) and relevancy/diversity measures.
— We enhance these algorithms with the direction awareness property.
— We propose new algorithms based on vertex selection (IL1, IL2, LM, $\gamma$-RLM) and query refinement (GSPARSE, FEED).
— We perform a rigorous set of experiments with various evaluation criteria and show that the proposed $\gamma$-RLM algorithm is suitable in practice for real-time diverse bibliographic search.

All of the algorithms in this paper are implemented and tested within the **advisor** and the best one ($\gamma$-RLM) will be used to power the system in a very near future.

## 2. BACKGROUND

### 2.1. Graph-based Citation Recommendation

Citation-analysis-based paper recommendation has been a popular problem since the '60s. There are methods that only take local neighbors (i.e., citations and references) into account, e.g., bibliographic coupling [Kessler 1963], cocitation [Small 1973], and CCIDF [Lawrence et al. 1999]. Recent studies, however, employ graph-based algorithms, such as Katz [Liben-Nowell and Kleinberg 2007], random walk with restarts [Pan et al. 2004], or well-known PageRank (PR) algorithm [Brin and Page 1998] to investigate the whole citation network. PaperRank [Gori and Pucci 2006], Ar-

---

ticleRank [Li and Willett 2009], and Katz distance-based methods [Strohman et al. 2007] are typical examples.

Ranking with Personalized PageRank (PPR) is a good way to find relevance scores of the papers. However, these algorithms treat the citations and references in the same way. This may not lead the researcher to recent and relevant papers if she is more interested in those. Old and well cited papers have an advantage with respect to the relevance scores since they usually have more edges in the graph. Hence the graph tends to have more and shorter paths from the seed papers to old papers. We previously defined the class of *direction aware* algorithms based on personalized PageRank, which can be tuned to reach a variety of citation patterns, allowing them to match the patterns of recent or traditional documents [Kucuktunc et al. 2012b]. We give the details of PageRank-based algorithms in Section 2.4.

## 2.2. Result Diversification on Graphs

The importance of diversity in ranking has been discussed in various data mining fields, including text retrieval [Carbonell and Goldstein 1998], recommender systems [Ziegler et al. 2005], online shopping [Vee et al. 2008], and web search [Clarke et al. 2008]. The topic is often addressed as a multi-objective optimization problem [Drosou and Pitoura 2010], which is shown to be NP-hard [Carterette 2009], and, therefore, some greedy [Agrawal et al. 2009; Haritsa 2009] and clustering-based [Liu and Jagadish 2009] heuristics were proposed. Although there is no single definition of diversity, different objective functions and axioms expected to be satisfied by a diversification system were discussed in [Gollapudi and Sharma 2009].

Diversification of the results of random walk-based methods on graphs only attracted attention recently. GRASSHOPPER is one of the earlier algorithms and addresses diversified ranking on graphs by vertex selection with absorbing random walks [Zhu et al. 2007]. It greedily selects the highest ranked vertex at each step and turns it into a sink for the next steps. Since the algorithm has a high time complexity, it is not scalable to large graphs. DIVRANK, on the other hand, combines the greedy vertex selection process in one unified step with the vertex reinforced random walk (VRRW) model [Mei et al. 2010]. This algorithm updates the transition matrix at each iteration with respect to the current or cumulative ranks of the nodes to introduce a *rich-gets-richer* mechanism to the ranking. But since the method updates the full transition matrix at each iteration, more iterations are needed for convergence; therefore, the computation cost increases. The shortcomings of those techniques were discussed in [Li and Yu 2011] in detail. [Tong et al. 2011] formalizes the problem from an optimization viewpoint, proposes the *goodness* measure to combine relevancy and diversity, and presents a near-optimal algorithm called DRAGON. These algorithms are further discussed in Section 3.

Coverage-based methods (such as the one in [Li and Yu 2011]) are also interesting for diversification purposes; however, they do not preserve the direction awareness property of the ranking function. Since our aim is to diversify the results of our paper recommendation service, we omitted the results of those coverage-based methods in our experiments.

## 2.3. Problem Definition

Let $G = (V, E)$ be a directed citation graph where $V = \{v_1, \ldots, v_n\}$ is the vertex set and $E$, the edge set, contains an edge $(u, v)$ if paper $u$ cites paper $v$. Let $\delta^+(u) = |\{(u, v) \in E\}|$ and $\delta^-(u) = |\{(v, u) \in E\}|$ be the number of references of and citations to paper $u$, respectively. We define the weight of an edge, $w(u, v)$, based on how important the citation is or how many times this paper is cited; however, for the sake of simplicity we take $w(u, v) = 1$ for all $(u, v) \in E$. Therefore, the nonsymmetric matrix $\mathbf{W} : V \times V$ becomes a 0-1 matrix. A summary of the notation used throughout the paper is given in Table 2.3.

We target the problem of paper recommendation assuming that the researcher has already collected a list of papers of interest [Kucuktunc et al. 2012b]. Therefore, the objective is to return papers that extend that list: given a set of $m$ seed papers $\mathcal{M} = \{p_1, \ldots, p_m\}$ s.t. $\mathcal{M} \subseteq V$, and a parameter $k$, return top-$k$ papers which are relevant to the ones in $\mathcal{M}$. With the diversity objective in mind, we want to recommend papers to be not only relevant to the query set $\mathcal{M}$, but also covering different topics around the query set.

Table I.
Notation

| | Symbol | Definition |
|---|---|---|
| **Graph** | $G = (V, E)$ | directed citation graph with $V$ nodes and $E$ edges |
| | $G' = (V, E')$ | undirected graph based on $G$ |
| | $n$ | $|V|$, number of vertices |
| | $\mathbf{W}$ | weights of the edges |
| | $w(u, v)$ | weight of the edge from $u$ to $v$ |
| | $\delta^-(v), \delta^+(v)$ | number of incoming or outgoing edges of $v$ |
| | $\delta(v)$ | $\delta^-(v) + \delta^+(v)$, number of neighbors of vertex $v$ |
| | $d(u, v)$ | shortest distance between $u$ and $v$ in $G$ |
| | $N_\ell(S)$ | $\ell$-step expansion set of $S \subseteq V$ |
| **Query** | $\mathcal{M}$ | a set of seed papers $\{p_1, \ldots, p_m\}$, $\mathcal{M} \subseteq V$ |
| | $m$ | $|\mathcal{M}|$, number of seed papers |
| | $k$ | required number of results, $k \leq n$ |
| | $R$ | a set of recommended vertices, $R \subseteq V$ and $|R| = k$ |
| | $d$ | damping factor of random walk with restart, $0 < d \leq 1$ |
| | $\kappa$ | direction-awareness parameter, $0 \leq \kappa \leq 1$ |
| **Random walk** | $p^*$ | prior distribution for personalized PageRank |
| | $t$ | iteration, or timestamp |
| | $\mathbf{p}_t$ | probability vector of being on a state at iteration $t$ |
| | $\eta_t$ | vector of number of visits at iteration $t$ |
| | $\mathbf{A}$ | structurally-symmetric $n \times n$ transition matrix based on $G$ |
| | $\mathbf{A}'$ | structurally-symmetric $n \times n$ transition matrix based on $G'$ |
| | $\mathbf{P}$ | $n \times n$ transition matrix for an iterative random walk |
| | $\pi$ | $\mathbf{p}_\infty$, stationary probability vector, $\sum_v \pi(v) = 1$ |
| | $\epsilon$ | convergence threshold |
| **Measures** | $S$ | a subset of vertices, $S = \{s_1, \ldots, s_k\}$, $S \subseteq V$ |
| | $\hat{S}$ | top-$k$ results, $\hat{S} = \arg\max_{S' \subseteq V, |S'| = k} \sum_{v \in S'} \pi_v$ |
| | $rel(S)$ | normalized relevance of the set |
| | $diff(S)$ | difference ratio of two sets |
| | $use(S)$ | usefulness of the set |
| | $dens_\ell(S)$ | $\ell$-step graph density |
| | $\sigma_\ell(S)$ | $\ell$-expansion ratio |

## 2.4. PageRank, Personalized PageRank, and direction-aware Personalized PageRank

Let $G' = (V, E')$ be an undirected graph of the citation graph, $p(u, v)$ be the transition probability between two nodes (states), and $d$ be the damping factor.

### 2.4.1. PageRank (PR). [Brin and Page 1998]

We can define a random walk on $G'$ arising from following the edges (links) with equal probability and a random restart at an arbitrary vertex with $(1 - d)$ teleportation probability. The probability distribution over the states follows the discrete time evolution equation:

$$\mathbf{p}_{t+1} = \mathbf{P}\,\mathbf{p}_t, \tag{1}$$

where $\mathbf{p}_t$ is the vector of probabilities of being on a certain state at iteration $t$, and $\mathbf{P}$ is the transition matrix defined as:

$$\mathbf{P}(u, v) = \begin{cases} (1-d)\frac{1}{n} + d\frac{1}{\delta(v)}, & \text{if } (u, v) \in E' \\ (1-d)\frac{1}{n}, & \text{otherwise.} \end{cases} \qquad (2)$$

If the network is ergodic (i.e., irreducible and non-periodic), Eq. 1 converges to a stationary distribution $\pi = \mathbf{P}\pi$ after a number of iterations. And the final distribution $\pi$ gives the PageRank scores of the nodes based on only *centrality*.

In practice the algorithm is said to have converged when the probability of the papers are stable, i.e., when the process is in a *steady state*. Let

$$\Delta_t = (\mathbf{p}_t(1) - \mathbf{p}_{t-1}(1), \dots, \mathbf{p}_t(n) - \mathbf{p}_{t-1}(n)) \qquad (3)$$

be the difference vector. We say that the process is in the steady state when the L2 norm of $\Delta_t$ is smaller than a given value $\epsilon$. That is,

$$\|\Delta_t\|_2 = \sqrt{\sum_{i \in V} (\mathbf{p}_t(i) - \mathbf{p}_{t-1}(i))^2} < \epsilon. \qquad (4)$$

### 2.4.2. Personalized PageRank (PPR). [Haveliwala 2002]

In our problem, a set of nodes $\mathcal{M}$ was given as a query, and we want the random walks to teleport to only those given nodes. Let us define a prior distribution $p^*$ such that:

$$p^*(u) = \begin{cases} 1/m, & \text{if } u \in \mathcal{M} \\ 0, & \text{otherwise.} \end{cases} \qquad (5)$$

If we substitute the two $(1/n)$s in Eq. 2 with $p^*$, we get a variant of PageRank, which is known as *personalized PageRank* or *topic-sensitive PageRank* [Haveliwala 2002]. PaperRank [Gori and Pucci 2006] applies personalized PageRank method to the undirected citation graph $G'$.

### 2.4.3. Direction-aware Random Walk with Restart DARWR. [Kucuktunc et al. 2012b]

We defined a *direction awareness* parameter $\kappa \in [0, 1]$ to obtain more recent or traditional results in the top-$k$ documents [Kucuktunc et al. 2012b]. Given a query with inputs $k$, a *seed* paper set $\mathcal{M}$, damping factor $d$, and direction awareness parameter $\kappa$, Direction-aware Random Walk with Restart (DARWR) computes the steady-state probability vector $\pi$. The ranks of papers after iteration $t$ is computed with the following linear equation:

$$\mathbf{p}_{t+1} = p^* + \mathbf{A}\mathbf{p}_t, \qquad (6)$$

where $p^*$ is an $n \times 1$ restart probability vector calculated with (5), and $\mathbf{A}$ is a structurally-symmetric $n \times n$ matrix of edge weights, such that

$$a_{ij} = \begin{cases} \frac{d(1-\kappa)}{\delta^+(i)}, & \text{if } (i, j) \in E \\ \frac{d\kappa}{\delta^-(i)}, & \text{if } (j, i) \in E \\ 0, & \text{otherwise.} \end{cases} \qquad (7)$$

Note that $\mathbf{P}$ transition matrix of random walk-based methods is built using $\mathbf{A}$ and $p^*$; however, the edge weights in rows can be stored and read more efficiently with $\mathbf{A}$ in practice [Kucuktunc et al. 2012a].

Figure 2 shows that the direction-awareness parameter $\kappa$ can be adjusted to reach papers from different years with a range from late 1980's to 2010 for almost all values of $d$. In our online service, the parameter $\kappa$ can be set to a value of user's preference. It allows the user to obtain recent papers by setting $\kappa$ close to $1$ or finding older papers by setting $\kappa$ close to $0$.

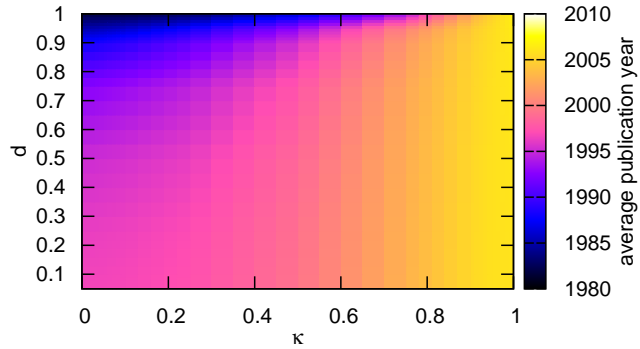

Fig. 2. Average publication year of top-10 recommendations by DARWR based on $d$ and $\kappa$.

## 3. DIVERSIFICATION METHODS

We classify the diversification methods for the paper recommendation problem based on whether the algorithm needs to rank the papers only once or multiple times. The first set of algorithms run a ranking function (e.g., PPR, DARWR, etc.) once and select a number of vertices to find a diverse result set. The algorithms in the second set run the ranking function $k$ times to select each result, and refine the search with some changes at each step. Although the former class of algorithms are preferred for practical use, they may not be able to reach to the intended diversity levels due to the highly greedy nature of the vertex selection process.

### 3.1. Diversification by vertex selection

The following approaches are used after getting the direction-aware relevancy (prestige) rankings of the vertices for a given set of seed nodes. The ranking function is selected as DARWR with parameters $(\kappa, d)$.

#### 3.1.1. DIVRANK: *Vertex-reinforced random walks.* [Mei et al. 2010]

For the random walk based methods mentioned so far, the probabilities in the transition matrix $\mathbf{P}$ do not change over the iterations. Using a variant of random walks, called *vertex-reinforced random walks* (VRRW) [Pemantle 1992], DIVRANK adjusts the transition matrix based on the number of visits to the vertices so far [Mei et al. 2010]. The original DIVRANK assumes that there is always an organic link for all the nodes returning back to the node itself which is followed with probability $(1 - \alpha)$:

$$p_0(u, v) = \begin{cases} \alpha \frac{w(u,v)}{\delta(i)}, & \text{if } u \neq v \\ 1 - \alpha, & \text{otherwise,} \end{cases} \tag{8}$$

where $w(u, v)$ is equal to $1$ for $(u, v) \in E'$, and $0$ otherwise. The transition matrix $\mathbf{P}_t$ at iteration $t$ is computed with

$$\mathbf{P}_t(u, v) = (1 - d) \, p^*(v) + d \, \frac{p_0(u, v) \, \eta_t(v)}{\sum_{z \in V} p_0(u, z) \, \eta_t(z)} \tag{9}$$

where $\eta_t(v)$ is the number of visits of vertex $v$. It ensures that the highly ranked nodes collect more value over the iterations, resulting in the so called *rich-gets-richer* mechanism.

In summary, for each iteration of the defined vertex-reinforced random walks, the transition probabilities from a vertex $u$ to its neighbors are adjusted by the number of times they are visited up to that point $\eta_t(v)$. Therefore, $u$ gives a high portion of its rank to the frequently visited neighbors. Since the tracking of $\eta_t$ is nontrivial, the authors propose to estimate it and provide two different estimation models. One way is to employ the *cumulative ranks* to estimate $\eta_t$ as

$$\mathbf{E}[\eta_t(v)] \propto \sum_{i=0}^{t} \mathbf{p}_i(v), \qquad (10)$$

and since the ranks will converge after sufficient number of iterations, it can also be estimated with *pointwise ranks*:

$$\mathbf{E}[\eta_t(v)] \propto \mathbf{p}_t(v). \qquad (11)$$

While adapting DIVRANK to our directional problem, we identified two problems: first, the initial ranks of all nodes should be set to a nonzero value; otherwise, the ranks cannot be distributed with Eq. 9 for both *pointwise* and *cumulative* estimation of $\eta_t$. Therefore, we set $p_0(v) = 1/n$ for all $v \in V$. Second, an organic link returning back to node itself enables the node to preserve its rank. This is problematic since $p^*$ is only set for seed papers, and they tend to get richer over time. However, our objective is to distribute the probabilities over other nodes to get a meaningful ranking. We solved this problem by removing the organic links of seed papers, hence distributing all of their ranks towards their neighbors instead of only $\alpha$ of them.

With the modifications above, we propose the *direction-aware* DIVRANK algorithm:

$$p_0(u,v) = \begin{cases} 0, & \text{if } u \in \mathcal{M}, u = v \\ \frac{(1-\kappa)}{\delta^+(u)}, & \text{if } u \in \mathcal{M}, u \neq v, (u,v) \in E \\ \frac{\kappa}{\delta^-(u)}, & \text{if } u \in \mathcal{M}, u \neq v, (v,u) \in E \\ (1-\alpha), & \text{if } u \notin \mathcal{M}, u = v \\ \alpha\frac{(1-\kappa)}{\delta^+(u)}, & \text{if } u \notin \mathcal{M}, u \neq v, (u,v) \in E \\ \alpha\frac{\kappa}{\delta^-(u)}, & \text{if } u \notin \mathcal{M}, u \neq v, (v,u) \in E \end{cases} \qquad (12)$$

where $\kappa$ is the direction awareness parameter. $p_0$ in Eq. 12 can be directly used in Eq. 9. Depending on the estimation method to be whether *cumulative* or *pointwise*, we refer to *direction-aware cumulative* DIVRANK as CDIVRANK, and *direction-aware pointwise* DIVRANK as PDIVRANK, respectively.

### 3.1.2. DRAGON: *Maximize the goodness measure.* [Tong et al. 2011]

One of many diversity/relevance optimization functions found in the literature is the *goodness* measure [Tong et al. 2011]. It is defined as:

$$f_{G'}(S) = 2\sum_{i \in S}\pi(i) - d\sum_{i,j \in S}\mathbf{A}'(j,i)\pi(j) - (1-d)\sum_{j \in S}\pi(j)\sum_{i \in S}p^*(i), \qquad (13)$$

where $\mathbf{A}'$ is the row-normalized adjacency matrix of the graph. The original algorithm runs on the undirected citation graph $G'$ and uses a greedy heuristic to find a near-optimal solution set. The direction-aware variant of the algorithm, running on the directed citation graph and using the ranking vector DARWR, is referred to as DRAGON. Accordingly, the direction-aware goodness measure $f_G(S)$ can be defined as:

$$f_G(S) = 2\sum_{i \in S}\pi(i) - d\kappa\sum_{i,j \in S}\mathbf{A}(j,i)\pi(j) - d(1-\kappa)\sum_{i,j \in S}\mathbf{A}(i,j)\pi(i), \qquad (14)$$

where $\kappa$ is the direction awareness parameter, $\mathbf{A}$ is the row-normalized adjacency matrix based on directed graph, and the last part of Eq. 13 is always zero $\left(\sum_{i \in S} p^*(i) = 0\right)$ since seed papers are never included in $S$.

### 3.1.3. IL1, IL2: *Ignore $\ell$-step expansion sets.*

Here we present a diversification approach that incrementally adds nodes to the recommendation list by ignoring their $\ell$-distance neighbors. In other words, after selecting the highest ranked node $r_1$ in the graph, the second highest ranked node $r_2$ is skipped if $r_2$ is in the expansion set of $r_1$. The *expansion set* with 1-distance neighbors is defined as $N(S) = S \cup \{v \in (V - S) : \exists u \in S, (u, v) \in E\}$, and the *$\ell$-step expansion set* is defined as

$$N_\ell(S) = S \cup \{v \in (V - S) : \exists u \in S, d(u, v) \leq \ell\}, \tag{15}$$

in [Li and Yu 2011]

Based on the parameter $\ell$, the results do not include direct references or citations of another recommendation ($\ell = 1$, referred to as IL1), and the papers that can be suggested with cocitation or cocoupling methods ($\ell = 2$, referred to as IL2).

### 3.1.4. LM: *Choose local maximas.*

Because of the smoothing process of random walks, frequently visited nodes tend to increase the ranks of its neighbors [Mei et al. 2010]. Therefore, we argue that computing local maximas and returning top-$k$ of them will guarantee that the nodes returned this way are recommended by taking the smoothing process of random walks into account.

Once the ranks are computed, the straightforward approach for getting the local maximas is to iterate over each node in the graph and check if its rank is greater than all of its neighbors' with a $\mathcal{O}(|E|)$ algorithm. However, the algorithm runs much faster in practice since every rank comparison between two unmarked nodes (either local maxima or not) will mark one of them. The LM algorithm is given in Alg. 1.

---

**ALGORITHM 1:** Diversify with local maximas (LM)

---

**Input**: $G' = (V, E')$: an undirected citation graph
$\pi$: ranks or stationary probabilities of the nodes in $V$
$k$: required number of recommendations
**Output**: A list of recommendations $S$
$L \leftarrow$ empty list of $(v, \pi_v)$
**for each** $v \in V$ **do**
    $lm[v] \leftarrow$ LocalMax
**for each** $v \in V$ **do**
    **if** $lm[v] =$ LocalMax **then**
       **for each** $v' \in adj[v]$ **do**
          **if** $\pi_{v'} < \pi_v$ **then**
             $lm[v'] \leftarrow$ NotLocalMax
          **else**
             $lm[v] \leftarrow$ NotLocalMax
             **break**
       **if** $lm[v] =$ LocalMax **then**
          $L \leftarrow L \cup \{(v, \pi_v)\}$
SORT$(L)$ w.r.t $\pi_i$ non-increasing
$S \leftarrow L[1..k].v$, i.e., top-$k$ vertices
**return** $S$

---

*3.1.5. $\gamma$-RLM: Choose relaxed local maximas.*

The drawback of diversifying with local maximas is that for large $k$'s (i.e., $k > 10$), the results of the recommendation algorithm are generally no longer related to the queried seed papers, but some popular ones in unrelated fields, e.g., a set of well-cited physics papers can be returned for a computer science related query. Although this might improve the diversity, it hurts the relevancy, hence, the results will no longer useful to the user.

In order to keep the results within reasonable relevancy to the query and the diversify them, we relax the algorithm by incrementally getting local maximas within top-$\gamma k$ results until $|S| = k$, and removing the selected vertices from the subgraph for the next local maxima selection. We refer this algorithm to as parameterized relaxed local maxima ($\gamma$-RLM). Note that 1-RLM reduces to DARWR and $\infty$-RLM reduces to LM. The outline of the algorithm is given in Alg. 2. In the experiments, we select $\gamma = k$ and refer this algorithm as $k$-RLM. Furthermore, we devise another experiment to see the effects of $\gamma$ with respect to different measures.

---

**ALGORITHM 2:** Diversify with relaxed local maximas ($\gamma$-RLM)

---

**Input**: $G' = (V, E')$: an undirected citation graph
$\pi$: ranks or stationary probabilities of the nodes in $V$
$k$: required number of recommendations
$\gamma$: relaxation parameter
**Output**: A list of recommendations $S$
$T \leftarrow \text{SORT}(V)$ w.r.t. $\pi_i$ non-increasing
$R \leftarrow T[1 : \gamma k]$
**while** $|S| < k$ **do**
    $R' \leftarrow \text{FINDLOCALMAXIMAS}(G, R, \pi)$
    **if** $|R'| > k - |S|$ **then**
        $\text{SORT}(R')$ w.r.t. $\pi_i$ non-increasing
        $R' \leftarrow R'[1 : (k - |S|)]$
    $S \leftarrow S \cup R'$
    $R \leftarrow R \setminus R'$
**return** $S$

---

## 3.2. Diversification by query refinement

In this set of diversification algorithms, the ranking function is called multiple times while some of the parameters or graph structure are altered between those rankings.

*3.2.1. GRASSHOPPER: Incremental ranking using absorbing random walks.* [Zhu et al. 2007]

GRASSHOPPER [Zhu et al. 2007] is a well-known diversification algorithm which ranks the graph multiple times by turning at each iteration the highest-ranked vertex into a sink node (A sink node only has a single outgoing edge to itself, so that all its rank stays trapped within the sink). Since the probabilities will be collected by the sink vertices when the random walk converges, the method estimates the ranks with the number of visits to each node before convergence.

The original method uses a matrix inversion to find the expected number of visits; however, inverting a sparse matrix makes it dense, which is not practical for the large and sparse citation graph we are using. Therefore, we estimate the number of visits by iteratively computing the cumulative ranks of the nodes with DARWR.

### 3.2.2. GSPARSE: *Incremental ranking by graph sparsification.*

In this algorithm, in contrast with GRASSHOPPER, after executing the ranking function, we propose to sparsify the graph by removing all reference and citation edges around the highest ranked node and repeat the process until $k$ nodes are selected in total. Note that GRASSHOPPER converts the selected node into a sink node while GSPARSE disconnects it from the graph (see Alg. 3 for details). This way, the graph around that node becomes less dense, hence, the nodes will attract less visits in a random walk.

---

**ALGORITHM 3:** Diversify by graph sparsification (GSPARSE)

---

**Input**: $G = (V, E)$: a directed citation graph
$\mathcal{M}$: query, a set of seed nodes
$k$: required number of recommendations
**Output**: A list of recommendations $S$
$S \leftarrow []$
$G' \leftarrow G$
**for** $iter = 1 \rightarrow k$ **do**
$\quad$ $ranks \leftarrow \text{DARWR}(G' = (V', E'), \mathcal{M})$
$\quad$ $v \leftarrow \arg\max(ranks)$
$\quad$ $S \leftarrow S \cup \{v\}$
$\quad$ **for each** $v' \in adj[v]$ **do**
$\quad\quad$ $E' \leftarrow E' \setminus \{(v, v')\}$
$\quad$ $V' \leftarrow V' \setminus \{v\}$
**return** $S$

---

### 3.2.3. FEED: *Feedback based on graph distance.*

For all the incremental algorithms, once the ranking function returns the most relevant node, it is most likely closer to some seed nodes than to other ones. In the next step, to obtain a different recommendation, one can decrease the importance of the closest seed nodes and increase the importance of the farthest seed nodes.

Following this idea, we argue that the prior probability vector $p^*$ can be adjusted with the inverse of the graph distance between the results and the seed papers. Let $p_t^*$ be the prior probability of the DARWR algorithm at step $t \leq k$. It is initialized as

$$p_0^*(v) = \begin{cases} 0, & \text{if } v \notin \mathcal{M}, \\ \frac{1}{|\mathcal{M}|}, & \text{if } v \in \mathcal{M}, \end{cases} \qquad (16)$$

and computed for the next iterations with

$$p_t^*(v) = \begin{cases} 0, & \text{if } v \notin \mathcal{M}, \\ \frac{median(\{d(r_0, v), ..., d(r_{t-1}, v)\})}{\sum_{v' \in \mathcal{M}} median(\{d(r_0, v'), ..., d(r_{t-1}, v')\})} & otherwise, \end{cases} \qquad (17)$$

where $r_i$ is the node selected at iteration $0 \leq i \leq t - 1$, and $d(u, v)$ is the shortest distance between $u$ and $v$. With an efficient implementation, computing $median(\{d(r_0, v), \ldots, d(r_{t-1}, v)\})$ is made fast by pre-computing the distances between all seed nodes and all vertices in the graph. If $\min$ was used when a node is selected near a seed node, the function would always return 1, which is undesired. On the other hand, if $\text{avg}$ was used when the seed nodes are far away from each other, a large distance between a selected and a sink node would affect the process for many iterations. $median$ is preferred over these functions because it gives meaningful values for the mentioned cases.

## 4. EXPERIMENTS

### 4.1. Evaluation measures

The relevancy and diversity of the results should be measured with separate methods since the problem is multi-criteria. For both diversity and relevancy parts, we evaluate the quality of the results with a number of measures.

#### 4.1.1. Relevancy measures.

**Normalized relevance:** The relevancy score of a set can be computed by comparing the original ranking scores of the resulting set with the top-$k$ ranking list [Tong et al. 2011], defined as

$$rel(S) = \frac{\sum_{v \in S} \pi_v}{\sum_{i=1}^{k} \hat{\pi}_i},$$ (18)

where $\hat{\pi}$ is the sorted ranks in non-increasing order.

**Difference ratio:** Results of a diversity method is expected to be somewhat different than the top-$k$ relevant set of results since, as our experiments will show, the set of nodes recommended by the original DARWR are not diverse enough. This is expected since highly ranked nodes will also increase the ranks of their neighbors [Mei et al. 2010]. Nevertheless, the original result set has the utmost relevancy. This fact can mislead the evaluation of the experimental results. Therefore, we decided to measure the difference of each result set from the set of original top-$k$ nodes. The difference ratio is computed with

$$diff(S, \hat{S}) = 1 - \frac{|S \cap \hat{S}|}{|S|},$$ (19)

where $\hat{S}$ is the top-$k$ relevant set.

**Usefulness:** The original ranking scores $\pi$ actually show the usefulness of the nodes. Since these scores usually follow a power law distribution, the high ranked nodes collect most of the scores and the contribution of two low-ranked nodes to the *rel* measure can be almost the same even though the gap between their positions in the ranking is huge. Yet, the one with the slightly higher score might be useful where the other might not due to this gap. We propose the *usefulness* metric to capture what percentage of the results are actually *useful* regarding their position in the ranking:

$$use(S) = \frac{|\{v \in S : \pi_v \leq \tilde{\pi}\}|}{|S|},$$ (20)

where $\tilde{\pi} = \hat{\pi}_{10 \times k}$, i.e., the relevancy score of the node with rank $10 \times k$, for $k = |S|$, and $use(S)$ gives the ratio of the recommendations that are within top $10 \times k$ of the relevancy list.

#### 4.1.2. Diversity measures.

**$\ell$-step graph density:** A variant of graph density measure is the $\ell$-step graph density [Tong et al. 2011], which takes the effect of in-direct neighbors into account. It is computed with

$$dens_\ell(S) = \frac{\sum_{u,v \in S, u \neq v} d_\ell(u, v)}{|S| \times (|S| - 1)},$$ (21)

where $d_\ell(u, v) = 1$ when $v$ is reachable from $u$ within $\ell$ steps, i.e., $d(u, v) \leq \ell$, and $0$ otherwise. The inverse of $D_\ell(S)$ is used for the evaluation of diversity in [Mei et al. 2010].

**$\ell$-expansion ratio:** Other diversity measures, *expansion ratio* and its variant $\ell$-*expansion ratio* [Li and Yu 2011] measure the coverage of the graph by the solution set. They are computed using the $\ell$-step expansion set given in Eq. 15 as

$$\sigma_\ell(S) = \frac{|N_\ell(S)|}{n}. \tag{22}$$

### 4.1.3. Other criteria.
**Goodness:** Given in Eq. 14.

**Average year:** The average publication year of the recommendation set.

**Average pairwise distance:** Pairwise shortest distances between the recommendations is a measure of how connected or distant the recommendations are to each other. It is computed with

$$AVG\_pairwise\_dist\,(S) = \frac{\sum_{u,v \in S, u \neq v} d(u,v)}{|S| \times (|S|-1)}. \tag{23}$$

**Average MIN distance to $\mathcal{M}$:** Distance of the recommendations to the closest seed paper is a measure of relevance regarding the query:

$$AVG\_min\_dist\_to\_\mathcal{M}(S) = \frac{\sum_{v \in S} \min_{p \in \mathcal{M}} d(s,p)}{|S|}. \tag{24}$$

## 4.2. Dataset collection and queries

We retrieved information on 1.9M computer science articles (as of March 2012) from DBLP [Ley 2009], 740K technical reports on physics, mathematics, and computer science from arXiv[3], and 40K publications from HAL-Inria[4] open access library. This data is well-formatted and disambiguated; however, it contains very few citation information (less than 470K edges). CiteSeer[5] is used to increase the number of paper-to-paper relations of computer science publications, but most of its data are automatically generated [Giles et al. 1998] and are often erroneous. We mapped each document in CiteSeer to at most one document in each dataset with the title information (using an inverted index on title words and Levenshtein distance) and publication years. Using the disjoint sets, we merged the papers and their corresponding metadata from four datasets. The papers without any references or incoming citations are discarded. The final citation graph has about 1M papers and 6M references, and is currently being used in our service.

The query set is composed of the actual queries submitted to the **advisor** service. We selected about 240 queries where each query is a set $\mathcal{M}$ of paper ids obtained from the bibliography files submitted by the users of the service who agreed to donating their queries for research purposes. $|\mathcal{M}|$ varies between 1 and 130, with an average of 24.35.

## 4.3. Results

We run the algorithms on the **advisor** citation graph with varying $k$ values (i.e., $k \in \{5, 10, 20, 50, 100\}$) and with the following parameters: $\alpha$ in VRRW (see Eq. 8) is selected as $0.25$ as suggested by the authors. For the DARWR ranking, we use the default settings of the service, which are $d = 0.9$ for damping factor, and $\kappa = 0.75$ to get more recommendations from recent publications. In each run, the selected algorithm gives a set of recommendations $S$, where $S \subseteq V$, $|S| = k$, and $S \cap \mathcal{M} = \emptyset$. The relevancy and diversity measures are computed on $S$, unless specified otherwise, and the average of each measure is displayed for different $k$ values. The standard deviations are omitted from the plots since they are negligible.

---

The strategy that we choose to select the best algorithm for our purpose (i.e., diversification of the results of the **advisor** recommendations) is to eliminate the algorithms one-by-one with respect to their results on various relevancy and diversity measures. This approach might sound quite unorthodox; however, it is probably the best way since (1) it is not clear if scoring extremely high or low is better for some of the measures (e.g., normalized relevancy scores of $0$ and $1$ are not preferred for a diversification method since they are the results of random and top-$k$ algorithm, respectively), and (2) there was no method that performed the best in all metrics. Although we eliminate some algorithms, for completeness, we give their results for all the measures.

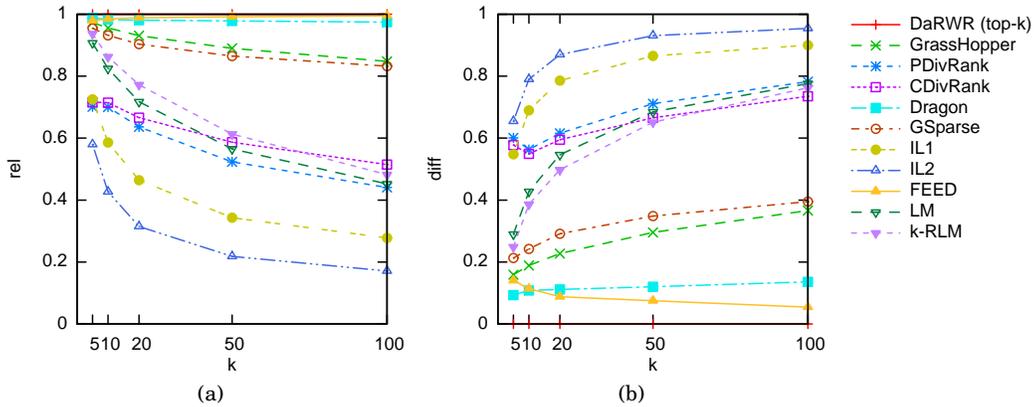

Fig. 3.   Normalized relevance (a) and difference ratio (b) of the result set with respect to top-$k$ results. Note that $rel_{DARWR} = 1$ and $diff_{DARWR} = 0$ since we compare the result set against itself.

Fig. 3 shows the normalized relevancy and difference ratio of the recommendations compared to top-$k$ results. It is arguable that a diversity-intended algorithm should maximize the relevancy since top-$k$ results will always get the highest score, yet those have almost no value considering the diversity part. On the other hand, having a very low relevancy score would tell us that the vertices are selected randomly, having no connection to the query at all.

Since the normalized relevancy does not give us a clear idea of what is expected from those diversity-intended methods, we compare the set difference of the results from top-$k$ relevant recommendations. Fig. 3-b clearly shows that two methods, namely FEED and DRAGON, give result sets that are only 10-15% different than the top-$k$. In other words, the results of FEED and DRAGON differ in only one element when $k = 10$. The experiments show that high difference ratio and low $rel$ scores of IL1 and IL2 can be problematic in practice.

Next, we evaluate the algorithms with respect to their usefulness scores. This experiment shows clearly that IL2 has a very low usefulness compared to other algorithms, scoring less than 50% for $k \geq 10$, meaning that more than half of its recommendations are out of top-$10k$ range (see Fig 4-a). DRAGON, FEED, and the original top-$k$ results score well on direction-aware goodness (Fig 4-b); however, this also means that the goodness measure gives more importance to relevancy and little importance to diversity.

Graph density is frequently used as a diversity measure in the literature [Tong et al. 2011; Li and Yu 2011]. IL1 and IL2 minimizes the $\ell$-step graph density by construction

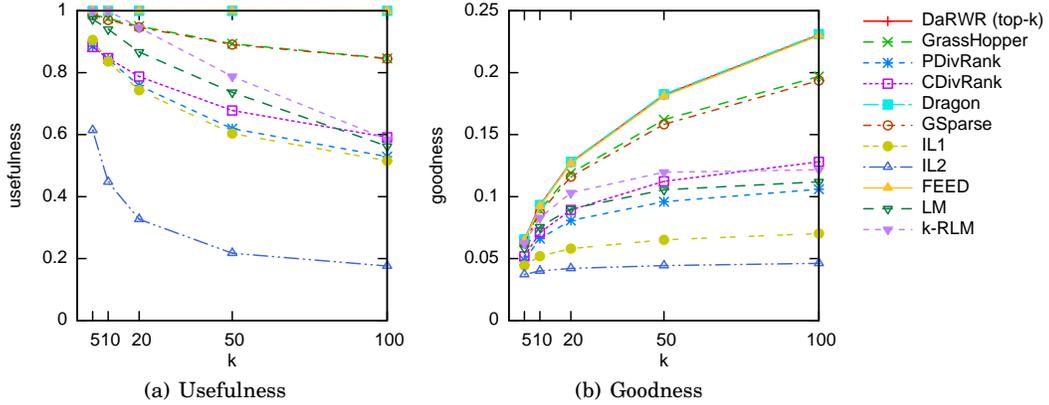

Fig. 4. Scores based on usefulness (a) and goodness (b) measures. Note that the results of DARWR are similar to hence, hidden behind the results of FEED and DRAGON in these plots.

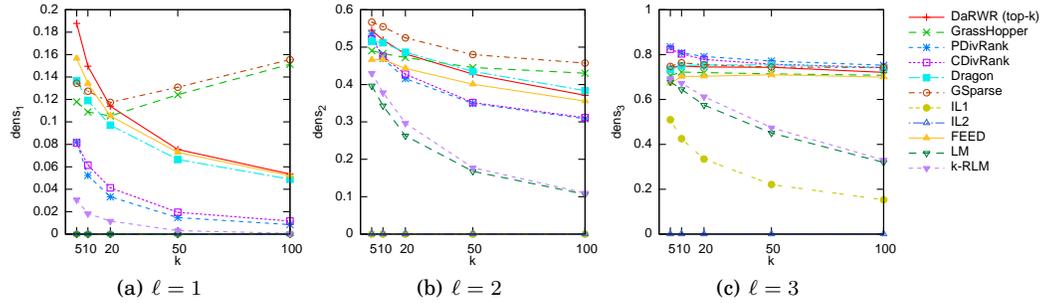

Fig. 5. $\ell$-step graph density of the results. Note that *dens* is 0 for IL1,IL2,LM at $\ell = 1$, IL1,IL2 at $\ell = 2$, and IL2 at $\ell = 3$.

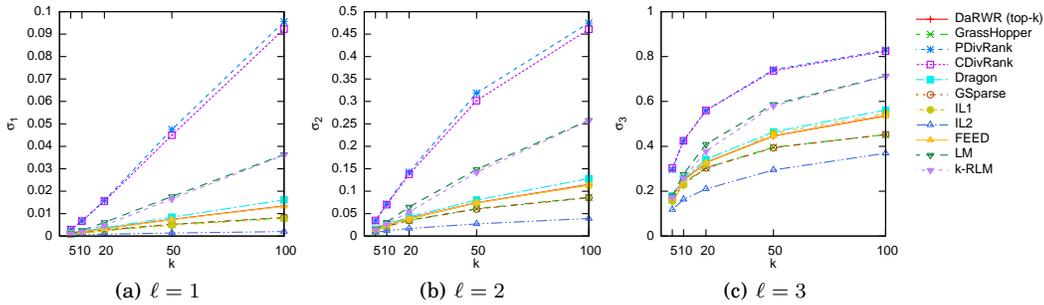

Fig. 6. $\ell$-step expansion ratio of the results.

(see Fig 5); so it would not be fair to compare other methods against these. LM, $k$-RLM, and DIVRANK variants, on the other hand, seem very promising for such a diversity objective. The same algorithms also perform good on $\ell$-step expansion ratio (see Fig 6), which is related to the coverage of the graph with the recommendations. GRASSHOPPER and GSPARSE perform convincingly worse in these diversity metrics. In particular, they are more dense than the results of DARWR.

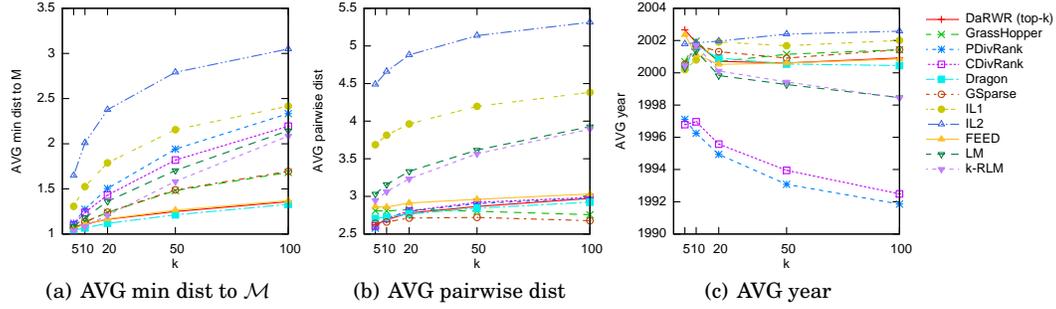

Fig. 7. Results based on average minimum distance to the query (a), average pairwise shortest distance between the recommended papers (b), and average publication year (c).

After evaluating the results on various relevancy and diversity metrics, we are left with only a couple of methods that performed well on almost all of the measures: LM, $k$-RLM, and DIVRANK variants. In Fig. 7, however, we observe that PDIVRANK and CDIVRANK methods give a set of results that are more connected (i.e., have a low average pairwise distance, see Fig. 7-b) and do not recommend recent publications (see Fig. 7-c) although the $\kappa$ parameter is set accordingly. Since we are searching for an effective diversification method that runs on top of DARWR, DIVRANK variants are no longer good candidates.

## 4.4. Efficiency

Running time of the algorithms is also crucial for the web service since all the recommendations are computed in real-time. We run the experiments on the same cluster that the service is currently using. It has a 2.4GHz *AMD Opteron* CPU and 4GB of main memory. The CPU has 64KB L1 and 1MB L2 caches. DARWR method and the dataset are also optimized based on the techniques given in [Kucuktunc et al. 2012a].

It was expected that the complexity of the methods based on query refinement depend on and increase linearly with $k$. As seen in Fig 8, GRASSHOPPER, GSPARSE, and FEED methods have the longest runtimes, even though they were faster than DI-VRANK variants for $k \leq 10$. This behavior was also mentioned in [Mei et al. 2010]. The running time of DRAGON is slightly higher than LM and $k$-RLM since it updates the goodness vector after finding each result.

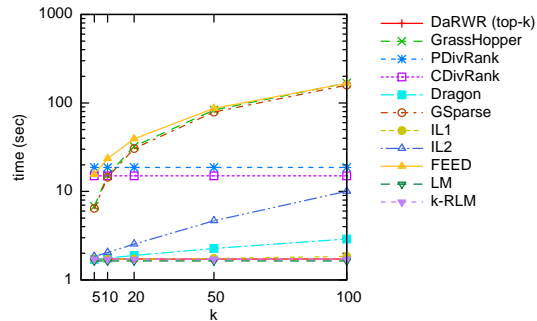

Fig. 8. Running times of the algorithms for varying $k$. Note that the running times of DARWR, IL1, IL2, LM, and $k$-RLM are less than the others and very close to each other.

In short, query refinement-based methods (GRASSHOPPER, GSPARSE, FEED) have linearly increasing runtimes. DIVRANK variants (PDIVRANK, CDIVRANK) requires more iterations, therefore, more time to converge. Finally, DRAGON, and especially LM and $k$-RLM are extremely efficient compared to other methods.

### 4.5. Selecting the best method for the service

Our experiments on different relevancy and diversity measures show that

— FEED and DRAGON return almost the same result set as top-$k$, while the graph density and expansion ratio measures also imply low diversity for their results,
— the results of IL2 have a very low usefulness,
— the results of IL1 have a low relevancy and high difference ratio,
— GRASSHOPPER and GSPARSE perform worse based on the diversity measures, and
— DIVRANK variants sacrifice direction-awareness for the sake of diversity.

On the other hand, LM and $k$-RLM methods perform convincingly good in almost all experiments, and have a better running time compared to others. $k$-RLM is slightly better than LM since it also improves the relevancy of the set to the query. We display the results of $\gamma$-RLM with varying $\gamma$ parameters in Figure 9.

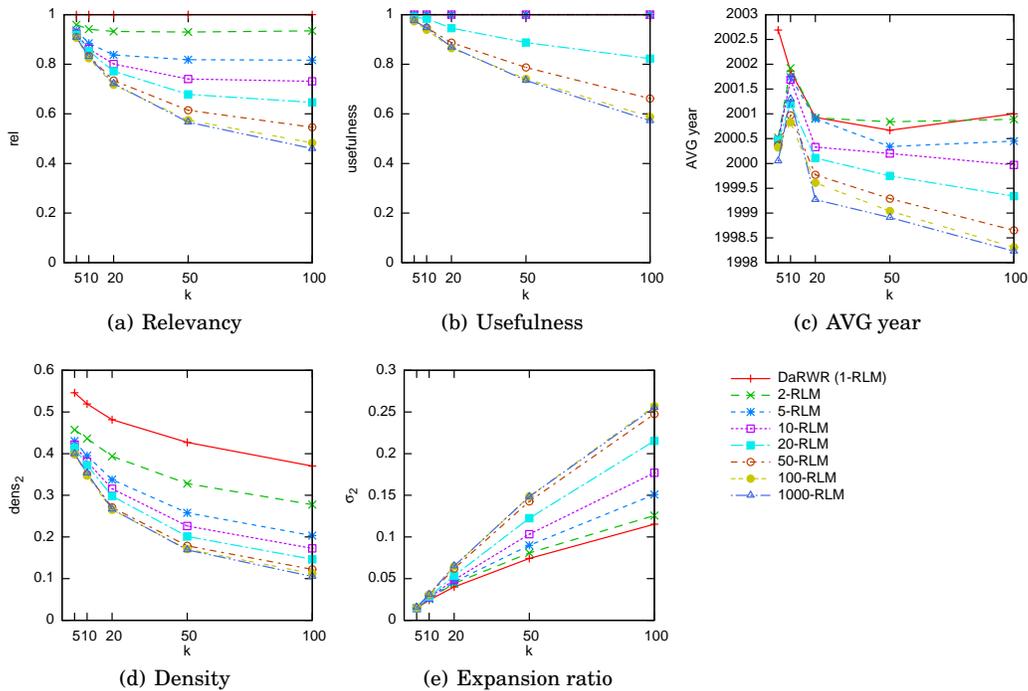

Fig. 9. Results of $\gamma$-RLM with varying parameters.

The evaluations on different metrics show that $\gamma$-RLM is able to sweep through the search space between all relevant (results of DARWR) and all diverse (results of LM) with a varying $\gamma$ parameter. Therefore, this parameter can be set depending on the data and/or diversity requirements of the application.

## 5. CONCLUSIONS

In this work, we addressed the diversification of paper recommendations of the **advisor** service, which ranks papers in the literature with a direction-aware personalized PageRank algorithm. While giving a survey of diversity methods designed specifically for random walk-based rankings, we adapted those methods to our direction-aware problem, and proposed some new ones based on vertex selection and query refinement. Our experiments with various relevancy and diversity measures show that the proposed $\gamma$-RLM algorithm can be preferred for both its efficiency and effectiveness.

We also learnt from our experiments that if one relevancy and one diversity measure was selected to evaluate the results of a diversification method –which is the case for many studies in this field–, a randomized algorithm that returns some of top ranked (relevant) results as well as some other random results will maximize those two measures, even though the output would be far from satisfactory for the user. Therefore, the results of such diversification algorithms should be examined with respect to multiple relevancy, coverage, and difference measures. We believe that we were able to take steps towards a better evaluation of diversity methods in this paper. Furthermore, our arguments and conclusions are also valid for cases without direction-awareness requirement. As a future work, we will investigate the techniques used in this paper for other applications and discuss how to improve the graph-based diversification algorithms as well as the evaluation methods.

## REFERENCES

AGRAWAL, R., GOLLAPUDI, S., HALVERSON, A., AND IEONG, S. 2009. Diversifying search results. In *Proc. ACM Int'l Conf. Web Search and Data Mining*. 5–14.

BRIN, S. AND PAGE, L. 1998. The anatomy of a large-scale hypertextual web search engine. In *Proc. Int'l Conf. World Wide Web*.

CARBONELL, J. AND GOLDSTEIN, J. 1998. The use of MMR, diversity-based reranking for reordering documents and producing summaries. In *Proc. Int'l ACM SIGIR Conf. Research and Development in Information Retrieval*. 335–336.

CARTERETTE, B. 2009. An analysis of NP-completeness in novelty and diversity ranking. In *Proc. Int'l Conf. Theory of Information Retrieval*. 200–211.

CLARKE, C. L., KOLLA, M., CORMACK, G. V., VECHTOMOVA, O., ASHKAN, A., BÜTTCHER, S., AND MACKINNON, I. 2008. Novelty and diversity in information retrieval evaluation. In *Proc. Int'l ACM SIGIR Conf. Research and Development in Information Retrieval*. 659–666.

DROSOU, M. AND PITOURA, E. 2010. Search result diversification. *SIGMOD Rec. 39*, 1, 41–47.

GILES, C. L., BOLLACKER, K. D., AND LAWRENCE, S. 1998. Citeseer: An automatic citation indexing system. In *Proc. ACM Conf. Digital Libraries*.

GOLLAPUDI, S. AND SHARMA, A. 2009. An axiomatic approach for result diversification. In *Proc. Int'l Conf. World Wide Web*. 381–390.

GORI, M. AND PUCCI, A. 2006. Research paper recommender systems: A random-walk based approach. In *Proc. IEEE/WIC/ACM Web Intelligence*.

HARITSA, J. R. 2009. The KNDN problem: A quest for unity in diversity. *IEEE Data Eng. Bull. 32*, 4, 15–22.

HAVELIWALA, T. H. 2002. Topic-sensitive pagerank. In *Proc. Int'l Conf. World Wide Web*.

KESSLER, M. M. 1963. Bibliographic coupling between scientific papers. *American Documentation 14*, 10–25.

KUCUKTUNC, O., KAYA, K., SAULE, E., AND CATALYUREK, U. V. 2012a. Fast recommendation on bibliographic networks. In *Proc. IEEE/ACM Int'l Conf. Social Networks Analysis and Mining*.

KUCUKTUNC, O., SAULE, E., KAYA, K., AND CATALYUREK, U. V. 2012b. Direction awareness in citation recommendation. In *Proc. Int'l Workshop on Ranking in Databases (DBRank'12) in conjunction with VLDB'12*.

LAO, N. AND COHEN, W. 2010. Relational retrieval using a combination of path-constrained random walks. *Machine Learning 81*, 53–67.

LAWRENCE, S., GILES, C. L., AND BOLLACKER, K. 1999. Digital libraries and autonomous citation indexing. *Computer 32*, 67–71.

LEY, M. 2009. DBLP - some lessons learned. *Proc. VLDB Endowment 2*, 2, 1493–1500.

LI, J. AND WILLETT, P. 2009. ArticleRank: a PageRank-based alternative to numbers of citations for analyzing citation networks. *Proc. ASLIB 61*, 6.

LI, R.-H. AND YU, J. 2011. Scalable diversified ranking on large graphs. In *Proc. IEEE Int'l Conf. Data Mining*. 1152–1157.

LIANG, Y., LI, Q., AND QIAN, T. 2011. Finding relevant papers based on citation relations. In *Proc. Int'l Conf. Web-Age Information Management*.

LIBEN-NOWELL, D. AND KLEINBERG, J. M. 2007. The link-prediction problem for social networks. *JASIST 58*, 7, 1019–1031.

LIU, B. AND JAGADISH, H. V. 2009. Using trees to depict a forest. *Proc. VLDB Endowment 2*, 1, 133–144.

MA, N., GUAN, J., AND ZHAO, Y. 2008. Bringing pagerank to the citation analysis. *Inf. Process. Manage. 44*, 800–810.

MCNEE, S. M., ALBERT, I., COSLEY, D., GOPALKRISHNAN, P., LAM, S. K., RASHID, A. M., KONSTAN, J. A., AND RIEDL, J. 2002. On the recommending of citations for research papers. In *Proc. of ACM Computer Supported Cooperative Work*.

MEI, Q., GUO, J., AND RADEV, D. 2010. DivRank: the interplay of prestige and diversity in information networks. In *Proc. ACM SIGKDD Int'l Conf. Knowledge Discovery and Data Mining*.

PAN, J.-Y., YANG, H.-J., FALOUTSOS, C., AND DUYGULU, P. 2004. Automatic multimedia cross-modal correlation discovery. In *Proc. ACM Knowledge Discovery and Data Mining*.

PEMANTLE, R. 1992. Vertex-reinforced random walk. *Probab. Theory Related Fields 92*, 117–136.

PETERS, H. P. F., BRAAM, R. R., AND VAN RAAN, A. F. J. 1995. Cognitive resemblance and citation relations in chemical engineering publications. *Journal of the American Society for Information Science 46*, 1, 9–21.

SALTON, G. 1963. Associative document retrieval techniques using bibliographic information. *J. ACM 10*, 440–457.

SMALL, H. 1973. Co-citation in the scientific literature: A new measure of the relationship between two documents. *J. Am. Soc. Inf. Sci. 24*, 4, 265–269.

STROHMAN, T., CROFT, W. B., AND JENSEN, D. 2007. Recommending citations for academic papers. In *Proc. Int'l ACM SIGIR Conf. Research and Development in Information Retrieval*.

TONG, H., HE, J., WEN, Z., KONURU, R., AND LIN, C.-Y. 2011. Diversified ranking on large graphs: an optimization viewpoint. In *Proc. ACM SIGKDD Int'l Conf. Knowledge Discovery and Data Mining*. 1028–1036.

VEE, E., SRIVASTAVA, U., SHANMUGASUNDARAM, J., BHAT, P., AND AMER-YAHIA, S. 2008. Efficient computation of diverse query results. In *Proc. IEEE Int'l Conf. Data Engineering*. 228–236.

ZHU, X., GOLDBERG, A. B., GAEL, J. V., AND ANDRZEJEWSKI, D. 2007. Improving diversity in ranking using absorbing random walks. In *Proc. of HLT-NAACL*.

ZIEGLER, C.-N., MCNEE, S. M., KONSTAN, J. A., AND LAUSEN, G. 2005. Improving recommendation lists through topic diversification. In *Proc. Int'l Conf. World Wide Web*. 22–32.